%% file: main.tex
\documentclass{article}

\usepackage{cite} 
\usepackage{graphics}
\usepackage{graphicx}
\usepackage[english]{babel}
\usepackage{amsmath}
\usepackage{amsthm}
\usepackage[table,xcdraw]{xcolor}
\usepackage[font=footnotesize]{caption}
\usepackage{epsfig}
\usepackage{xfrac}
\usepackage{nicefrac}
\usepackage{psfrag}
\usepackage{times}
\usepackage{balance} 
\usepackage{amssymb}
\usepackage{amsfonts}
\usepackage{bm}
\usepackage[bottom]{footmisc}
\usepackage{mwe}
\usepackage{color}
\usepackage{booktabs,multirow}
\usepackage{tabularx}
\usepackage{tikz}
\usetikzlibrary{patterns,arrows}
\usepackage{physics}
\usepackage{verbatim}
\usepackage{subcaption}
\usepackage[detect-all]{siunitx}
\usepackage{hyperref}

\newcommand{\tblue}[1]{\textcolor{black}{#1}}

\newcommand{\eg}{{\sl e.g., }}
\newcommand{\ie}{{\sl i.e., }}
\newcommand{\BfPara}[1]{{\noindent\bf#1.}\xspace}

\definecolor{color1}{RGB}{228,26,28}
\definecolor{color2}{RGB}{55,126,184}
\definecolor{color3}{RGB}{77,175,74}
\definecolor{color4}{RGB}{152,78,163}
\definecolor{color5}{RGB}{255,127,0}
\definecolor{color6}{RGB}{200,200,200}

\usepackage{hyperref}

\usepackage[accepted]{icml2025}

\icmltitlerunning{Autonomous Base Station Deployment with Reinforcement Learning and Digital Network Twins}

\begin{document}

\onecolumn
\icmltitle{AutoBS: Autonomous Base Station Deployment with Reinforcement Learning and Digital Network Twins}




\begin{icmlauthorlist}
\icmlauthor{Ju-Hyung Lee}{nokia}
\icmlauthor{Andreas F. Molisch}{usc}
\end{icmlauthorlist}

\icmlaffiliation{nokia}{Nokia Technologies, Sunnyvale, CA, USA}
\icmlaffiliation{usc}{Department of Electrical and Computer Engineering, University of Southern California, Los Angeles, CA, USA}

\icmlcorrespondingauthor{Ju-Hyung Lee}{juhyung.lee@nokia.com}

\icmlkeywords{6G, Wireless Networks, Digital Network Twins, Base Station Deployment, Network Optimization, Ray Tracing, Reinforcement Learning, Proximal Policy Optimization}

\vskip 0.3in



\printAffiliationsAndNotice{}  % leave blank if no need to mention equal contribution

\input{section/abstract}
\input{section/sec_1}

\input{section/sec_2}

\input{section/sec_3}

\input{section/sec_4}

\input{section/conclusion}

% \vspace{.5em}
\newpage
{\bf Acknowledgements:} Part of this work was supported by Nokia Technologies and NSF-AoF grant 2133655. The authors gratefully acknowledge Arjun Balamwar and Yanqing Lu for their contributions to the framework design and simulations.
% \vspace{-1.em}
\bibliography{refs}
\bibliographystyle{icml2025}

%%%%%%%%%%%%%%%%%%%%%%%%%%%%%%%%%%%%%%%%%%%%%%%%%
%%%%%%%%%%%%%%%%%%%%%%%%%%%%%%%%%%%%%%%%%%%%%%%%%
% \newpage
\appendix
\onecolumn
\input{section/appendix}

%%%%%%%%%%%%%%%%%%%%%%%%%%%%%%%%%%%%%%%%%%%%%%%%%
%%%%%%%%%%%%%%%%%%%%%%%%%%%%%%%%%%%%%%%%%%%%%%%%%

\end{document}

%% file: section/abstract.tex
\begin{abstract}
This paper introduces \emph{AutoBS}, a reinforcement learning (RL)-based framework for optimal base station (BS) deployment in 6G radio access networks (RAN). AutoBS leverages the Proximal Policy Optimization (PPO) algorithm and fast, site-specific pathloss predictions from PMNet—a generative model for digital network twins (DNT). By efficiently learning deployment strategies that balance coverage and capacity, AutoBS achieves about 95\% of the capacity of exhaustive search in single BS scenarios (and in 90\% for multiple BSs), while cutting inference time from hours to milliseconds, making it highly suitable for real-time applications (\eg ad-hoc deployments). 
AutoBS therefore provides a scalable, automated solution for large-scale 6G networks, meeting the demands of dynamic environments with minimal computational overhead.
\end{abstract}

% AutoBS offers a scalable and automated solution for large-scale 6G networks, addressing the challenges of dynamic environments with minimal computational overhead.

%% file: section/sec_1.tex
\section{Introduction} \label{sec:intro}
The rollout of 6G networks demands higher base station (BS) density due to requirements for higher area spectral efficiency and the use of higher frequencies like millimeter-wave (mmWave). The latter offer enhanced bandwidth and low latency but suffer from severe signal attenuation and limited propagation range, particularly in complex urban environments. 
As a result, dense BS deployment becomes essential to maintain reliable coverage and capacity. Furthermore, cell-free massive MIMO, where a significant number of access points with one of a few antennas each are distributed over an area~\cite{interdonato2019ubiquitous} is anticipated to be widely used in 6G.  

However, optimizing BS placement in such environments presents significant challenges due to highly site-specific conditions.
Traditional BS deployment methods often rely on manual planning, based on real-world measurement of the propagation conditions~\cite{molisch2023wireless}, and/or computationally intensive ray-tracing (RT) simulations using tools like \emph{WirelessInsite} \cite{WirelessInsite} or SionnaRT \cite{sionna, aoudia2025sionnart}. 
The fact that these approaches are time- and labor- consuming makes them less suited for the dense deployment in 6G. Furthermore, they are not suited for real-time adaptation of network topology, \eg dynamic addition and placement of (mobile) BSs in reaction to change of user density, \eg at special events or for ad-hoc deployment by emergency responders or military applications. 

The network planning challenge can also be viewed within the framework of digital network twins (DNT), virtual replicas of physical environments that enable real-time simulation and optimization of network performance under site-specific conditions. Even within DNT frameworks, manually optimizing BS placement remains computationally prohibitive.

Hence, effective real-time optimization of large-scale networks requires the integration of machine learning (ML) approaches to automate and accelerate network planning, particularly for radio access network (RAN) deployments.

To address these challenges, we propose AutoBS—an autonomous BS deployment framework utilizing deep reinforcement learning (DRL) in conjunction with a generative DNT model for site-specific optimization. AutoBS integrates PMNet~\cite{lee2024scalable, lee2023robust}, a fast and accurate ML-based pathloss predictor, to efficiently compute rewards based on coverage and capacity metrics. By autonomously learning optimal BS placement strategies, AutoBS significantly reduces computational complexity and deployment time compared to conventional methods, making it highly suitable for real-time optimization in large-scale 6G RAN scenarios.

%%%%%%%%%%%%%%%%%%%%%%%%%%%%
\BfPara{Related Works}
Recent studies on DNT have focused primarily on developing realistic simulation platforms. \cite{pegurri2025a} integrated the \texttt{ns-3} network simulator with the differentiable ray tracer \texttt{SionnaRT}, creating a scalable full-stack DNT framework validated in dynamic urban vehicular scenarios\cite{pegurri2025b}. Additionally, learning-based physical-layer solutions have emerged, such as the neural multi-user MIMO receiver by \cite{cammerer2023a}, and the real-time neural receiver fine-tuned with site-specific propagation data by \cite{wiesmayr2024design}. In contrast to these works—centered around simulation fidelity or neural receiver design—\textbf{AutoBS} leverages a generative DNT model (PMNet) combined with a PPO agent to automate BS deployment decisions, shifting ML-driven DNT applications from device-level adaptations to network-level topology optimization.

\begin{figure}[!t]
    \centering
    \includegraphics[width=0.33\linewidth]{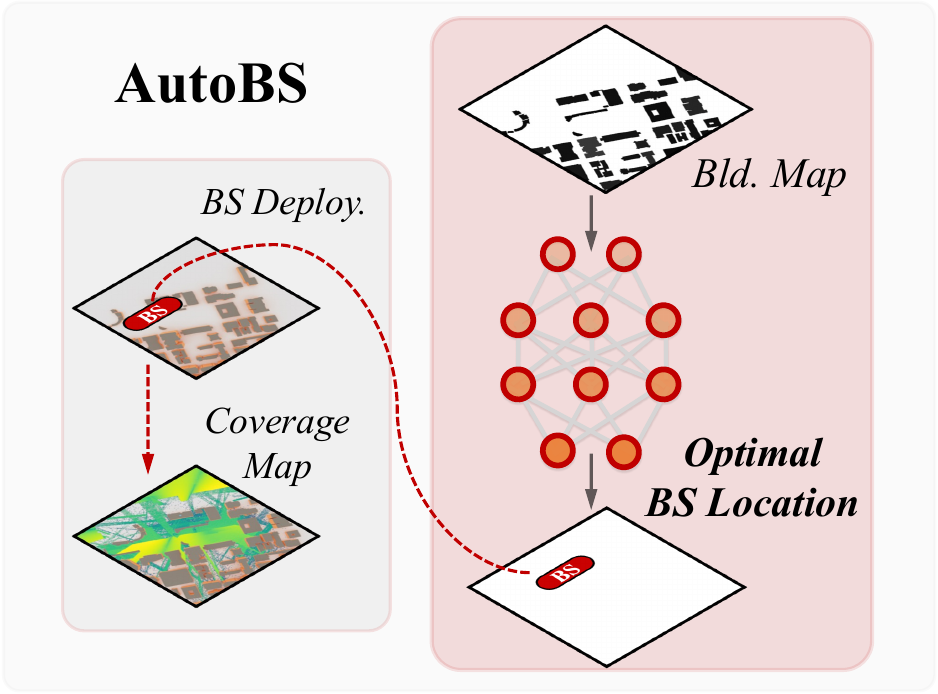}
    \caption{Overview of the AutoBS framework, where the DRL agent leverages PMNet’s generative pathloss map (DNT of the site) to evaluate coverage and determine optimal BS locations.}
    \label{fig:framework}
    \vspace{-1.em}
\end{figure}

%%%%%%%%%%%%%%%%%%%%%%%%%%%%
\BfPara{Contributions}
This paper presents AutoBS, an autonomous BS deployment framework utilizing DRL and a generative DNT model (\ie PMNet). Our key contributions are summarized as follows:
\begin{itemize}
\item \textbf{AutoBS Framework Design:} We introduce a novel DRL-based framework for BS deployment that incorporates PMNet for real-time, site-specific channel predictions\footnote{Selected results and codes are available at \href{https://github.com/abman23/autobs}{github.com/abman23/autobs}.}. 
As a generative model, PMNet enables fast, precise reward calculations, allowing AutoBS to effectively optimize coverage and capacity through an adaptive reward function (see \textbf{Fig. \ref{fig:training}} in Sec. \ref{sec:autobs}). Details on reward design are provided in \textbf{Table \ref{table:reward}} in Sec. \ref{sec:appendixa}.
    
\item \textbf{Support for Multi-BS Deployment:} AutoBS handles both static single BS and asynchronous multi-BS deployments, providing flexibility across diverse deployment scenarios (see \textbf{Fig. \ref{fig:comparison_coverage_multibs}} in Sec. \ref{sec:simulation}).

\item \textbf{Fast and Near-Optimal Deployment:} Simulations demonstrate that AutoBS reduces inference time from hours to milliseconds compared to exhaustive methods, particularly in multi-BS deployments, making it practical for real-time optimization (see \textbf{Table \ref{table:complexity}} in Sec. \ref{sec:simulation}).
\end{itemize}

%% file: section/sec_2.tex
\section{Methodology} \label{sec:method}

%%%%%%%%%%%%%%%%%%%%%%%%%%%%
\subsection{Problem Formulation and DRL Approach}
The BS deployment problem can be mathematically formulated as an optimization task. Let the BS location be represented by coordinates $(i, j)$ on a site-specific building map $m$. The objective function to optimize is:
\begin{align}
\max_{\{i,j\}} \sum_{m=1}^{M} \left( V_m + \nu C_m \right), \ \text{s.t.} \ \{i,j\} \in \mathcal{B}, \label{eq:OptObj}
\end{align}
where $V_m$ is the coverage, $C_m$ represents the capacity, and $\nu$ is a weighting factor that governs the trade-off between coverage and capacity. The set $B$ defines the allowable deployment locations for BS placement (such as rooftops or other designated sites within the environment).

AutoBS frames BS placement as a reinforcement learning task: a DRL agent interacts with the environment to learn optimal placement strategies. Through continuous reward feedback, the agent incrementally refines its decisions – capturing site-specific channel characteristics and improving network performance over time.

%%%%%%%%%%%%%%%%%%%%%%%%%%%%
\subsection{Reward Calculation via Digital Network Twins} \label{sec:method_2}
In DRL, efficient reward calculation is essential for guiding the learning process, where the agent explores numerous actions throughout training. However, calculating network performance metrics, such as coverage and capacity, based on BS placements—key components of the reward function—is computationally intensive. Traditional RT simulations, which can take tens of minutes to generate a single pathloss map, are impractical for real-time reinforcement learning. 

To overcome this challenge, AutoBS integrates \emph{PMNet} \cite{lee2024scalable}– a generative DNT model – for rapid reward computation. PMNet can predict a site-specific pathloss map within milliseconds (with an RMSE on the order of $10^{-2}$), enabling the PPO algorithm \cite{schulman2017proximal} to immediately evaluate network performance and obtain a reward after each deployment decision.
This rapid feedback loop allows the agent to efficiently simulate multi-BS deployment scenarios, significantly accelerating training by exploring many more placement options. In summary, PMNet enables fast and scalable reward evaluation, allowing AutoBS to operate in large, diverse network environments.

%% file: section/sec_3.tex
%%%%%%%%%%%%%%%%%%%%%%%%%%%%%%%%%%%%%%%%%%%%%%%%%
\section{AutoBS: An Autonomous Base Station Deployment Framework} \label{sec:autobs}

%%%%%%%%%%%%%%%%%%%%%%%%%%%%%%%%%%%%
\subsection{Architecture}

% The AutoBS framework automates optimal BS placement in 6G networks using DRL, with a PPO agent. PPO is well-suited for this task due to its balance between exploration (testing new deployment locations) and exploitation (refining placement strategies). By using a clipped objective function, PPO ensures stable policy updates and prevents large, destabilizing changes during training.
AutoBS uses a deep RL agent based on the PPO algorithm to automate optimal BS placement in 6G networks. PPO is well-suited to this task because it balances exploration (testing new deployment locations) and exploitation (refining placement strategies). By using a clipped objective function, PPO also ensures stable policy updates and prevents overly large, destabilizing changes during training.

The framework supports two deployment modes: (1) \emph{Static single BS deployment}, where one BS is optimally placed in a fixed environment; and (2) \emph{Asynchronous multi-BS deployment}, where multiple BSs are deployed asynchronously.

%%%%%%%%%%%%%%%%%%%%%%%%%%%%
\subsubsection{\textbf{Single (Static) BS Deployment}}
% In the \textit{static} single BS deployment scenario, the geographical environment and network conditions are assumed to be fixed. The input to the system is a site-specific map $\mathcal{S}$, which contains detailed information about buildings, terrain, and obstacles. The PPO agent processes this map and outputs the optimal location $(i, j)$ for placing the BS to maximize coverage and capacity. 
In the \emph{static} single BS scenario, the environment and network conditions remain fixed (time-invariant). The input to the agent is a site-specific map $\mathcal{S}$ containing details about buildings, terrain, and obstacles. The PPO agent processes this map and outputs the optimal coordinates $(i,j)$ for BS placement to maximize coverage and capacity.

% This approach is most effective for stable environments where network demand and user behavior remain relatively constant. Dynamic adjustments to network conditions are not required.

%%%%%%%%%%%%%%%%%%%%%%%%%%%%
\subsubsection{\textbf{Multi (Asynchronous) BS Deployment}}
In the \textit{asynchronous} multi-BS deployment scenario, BSs are deployed incrementally over time. After each deployment, the environment is updated to reflect the changes in network conditions, such as coverage, capacity, and user distribution. The PPO agent adjusts its strategy after each deployment based on real-time feedback, refining its decisions progressively. This scenario effectively models real-world situations in which a network requires gradual densification; the goal is to optimally determine locations for newly deployed BSs, while previously deployed BSs remain fixed.

% This approach allows the agent to adapt to dynamic environments where network conditions evolve. By deploying BSs one at a time, the agent can evaluate the impact of each BS placement, optimizing the network's performance iteratively. Asynchronous deployment is especially useful in practical network scenarios where external constraints, such as building leases or regulatory requirements, must be considered.

%%%%%%%%%%%%%%%%%%%%%%%%%%%%
\subsection{Evaluation Metrics}

AutoBS framework's performance is evaluated using two key metrics: \emph{Coverage} and \emph{Capacity}. These metrics are essential for assessing the quality of BS placements in terms of both spatial coverage and network throughput.

\BfPara{Coverage} represents the proportion of the area or users receiving sufficient signal strength, \eg received power, from the deployed BSs. It is calculated as the total area where the received power $P_{i,j}$ exceeds a predefined threshold $\mathrm{thr}$, ensuring satisfactory Quality of Service (QoS). The coverage $V$ is expressed as:
\begin{align}
V = \sum_{\{i,j\} \in \mathcal{R}}{v_{i,j}},
\end{align}
where $\mathcal{R}$ denotes the region where users (UEs) can be located, such as open areas without buildings, and $v_{i,j}$ represents a binary coverage indicator defined as: 
\begin{align}
v_{i,j} = \begin{cases}
    1, & P_{i,j} \geq \mathrm{thr}, \\
    0, & P_{i,j} < \mathrm{thr}.
\end{cases}
\end{align}
Here, the threshold $\mathrm{thr}$ is the minimum required received power expressed in linear units (Watts), typically corresponding to a value of approximately $-90$ [dBm]. This threshold value can vary depending on specific network QoS requirements, such as the minimum signal strength necessary for reliable communication.

\BfPara{Capacity} represents the maximum amount of data that can be transmitted or received over a network, which depends on the quality of the connection, \eg signal-to-noise ratio (SNR). The total capacity $C$ is given by:
\begin{align}
    C = \sum_{\{i,j\} \in \mathcal{R}} \log_2{\left( 1 + \text{SNR}_{i,j} \right)},
\end{align}
where $\text{SNR}_{i,j}$ is the signal-to-noise ratio at pixel $(i,j)$, calculated as:
\begin{equation}
    \text{SNR}_{i,j} = \frac{\sum_{k} P^{k}_{i,j}}{\sigma^2}.
\end{equation} 
Here, $P^{k}_{i,j}$ represents the received power from the $k$-th BS at pixel $(i,j)$, and $\sigma^2$ is the noise variance. In our simulations, we chose $\sigma^2 = \mathrm{thr} /4$, ensuring that the SNR at the coverage boundary is at least 6 dB.\footnote{In an actual DNT simulation, noise variance would obviously follow from the system bandwidth, transmit power, receiver noise figure, and antenna again; but since cell-edge SNR of 6 dB is a common planning goal, we directly used it here.} 

%%%%%%%%%%%%%%%%%%%%%%%%%%%%%%%%%%%%
\subsection{Training}
The AutoBS framework models the base station (BS) deployment problem as a Markov decision process (MDP), enabling the agent to interact with its environment to make optimal BS placement decisions.

%%%%%%%%%%%%%%%%%%%%%%%%%%%%
\subsubsection{MDP Design}
\BfPara{Environment}
\tblue{The environment for the PPO agent is defined by the interactions within a site-specific building map, where each deployed BS impacts the network performance metrics such as coverage and capacity. At each time step $t$, the agent interacts with this environment by observing a state $s_t \in \mathcal{S}$, which includes details on buildings, obstacles, and terrain specific to the deployment site. The agent then takes an action $a_t \in \mathcal{A}$, selecting a BS deployment location based on its learned policy $\pi$. After the action is executed, the environment transitions to a new state $s_{t+1}$, representing the updated network layout and performance with the new BS in place. The agent receives a reward $r_t$ based on the effectiveness of its decision, guiding it toward an optimal deployment policy $\pi^*$.}

\BfPara{State}
\tblue{The state $s_t$ at time step $t$ provides key information for decision-making about BS placement. Formally, the state is expressed as:
\begin{align}
s_t = \left\{ \mathcal{S} \right\},
\end{align}
where $\mathcal{S}$ represents the site-specific building map. This map encompasses critical details about building locations, obstacles, and terrain, all of which significantly influence signal propagation (\ie site-specific channel). 
}
% The minimal state design captures the key spatial characteristics required for optimal deployment while maintaining computational efficiency, enabling the agent to make well-informed placement decisions with minimal processing overhead.}

\BfPara{Action}
The action space $\mathcal{A}$ consists of potential BS deployment locations. At each time step $t$, the agent selects an action $a_t \in \mathcal{A}$, which corresponds to placing a BS at a specific geographical coordinate $(i, j)$ on the site-specific map. The action is represented as:
\begin{equation}
a_t = (i, j), \quad \{i,j\} \in \mathcal{B},
\end{equation}
where $\mathcal{B}$ is the set of all permissible deployment locations within the map. The agent deploys BSs sequentially, choosing new locations at each time step based on current network needs. This approach is computationally efficient, allowing the agent to adapt its strategy in real-time as it learns from previous decisions.

\BfPara{Reward}
The reward function $r_t$ incentivizes the agent to improve network performance by maximizing both coverage and capacity. The reward at time step $t$ is defined as:
\begin{equation}
r_t = \nu_1 V_{t} + \nu_2 C_{t} + \nu_3 P_{t},
\end{equation}
where $V_{t}$ and $C_{t}$ represent the improvements in coverage and capacity, respectively, and $P_{t}=\sum_{\{i,j\} \in \mathcal{R}} P_{i,j}$ denotes the total pathgain, for the $t$-th deployment. While the term $P_{t}$ is not explicitly included in the objective function in \eqref{eq:OptObj}, it strongly correlates with both coverage and capacity, providing a denser learning signal that accelerates training convergence. The weighting parameter $\nu$ adjusts the relative importance of coverage, capacity, and pathgain in the overall network optimization. For further details regarding the reward design, please refer to Appendix~\ref{sec:appendixa}. Note also that for more general DNT applications (\eg incorporating user demands), only the reward function needs to be adapted, while the remaining framework of AutoBS stays the same.

As mentioned in Sec. \ref{sec:method_2}, the reward must be recalculated after each BS deployment to reflect the updated network conditions. This typically requires coverage and capacity evaluation from pathloss map prediction, which are computationally expensive. To overcome this challenge, AutoBS framework integrates the  \emph{PMNet} model, which provides fast and accurate pathloss predictions, enabling efficient reward calculations. PMNet allows the agent to quickly assess the network performance after each BS placement, ensuring that the training process is both realistic and computationally feasible.

\begin{figure*}[!ht]
\centering
\includegraphics[width=.45\columnwidth]{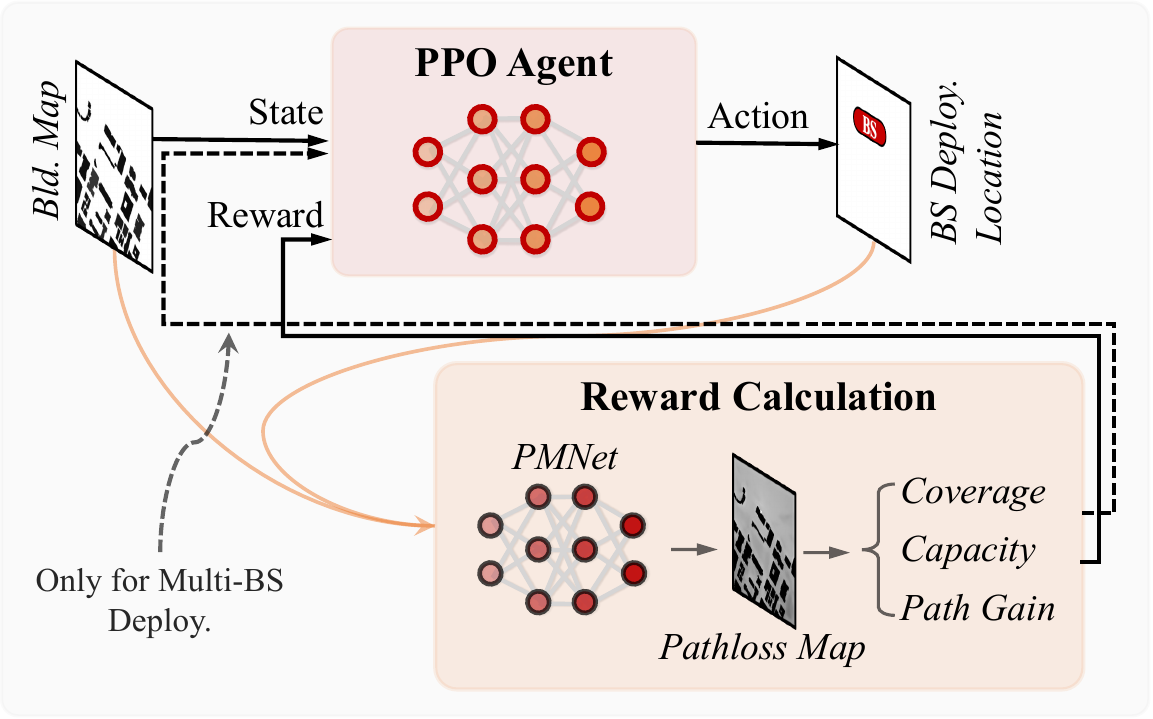}
\caption{Training process for AutoBS framework, with PMNet providing generative pathloss predictions to compute the reward at each step.}
\label{fig:training}
\vspace{-.5em}
\end{figure*} 
%%%%%%%%%%%%%%%%%%%%%%%%%%%%
\subsubsection{Training Process}

The PPO agent is trained using the PPO algorithm \cite{schulman2017proximal}, which allows the agent to iteratively interact with a simulated environment. The agent selects BS deployment actions and receives rewards based on improvements in network coverage and capacity, progressively refining its policy $\pi_\theta(a_t | s_t)$ over time.

At each time step $t$, the agent updates its policy to maximize the cumulative reward:
\begin{equation}
J(\theta) = \mathbb{E} \left[ \sum_{t=0}^{T} \gamma^t r_t \right],
\end{equation}
where $\gamma$ is the discount factor, $r_t$ is the reward received at time step $t$, and $T$ is the time horizon over which the agent aims to optimize its decisions. The discount factor $\gamma$ balances immediate and future rewards, promoting long-term planning in BS placement.

The PPO agent optimizes the following objective function:
\begin{equation}
\mathcal{L}(\theta) = \mathbb{E}_t \left[ \min \left( r_t(\theta) \hat{A}_t, \, \text{clip}(r_t(\theta), 1 - \epsilon, 1 + \epsilon) \hat{A}_t \right) \right],
\end{equation}
where $r_t(\theta)$ is the probability ratio between the updated and previous policies, $\epsilon$ is a clipping parameter to limit policy changes for stability, and $\hat{A}_t$ represents the advantage function, which quantifies how much better the current action is compared to the baseline.

For clarity and space reasons, we omit detailed descriptions of the PPO algorithm, and instead present a summary of the training process in Fig~\ref{fig:training}, which outlines the key steps involved in training the PPO agent.

%% file: section/sec_4.tex
%%%%%%%%%%%%%%%%%%%%%%%%%%%%%%%%%%%%%%%%%%%%%%%%%
\section{Experimental Results} \label{sec:simulation}

\subsection{Simulation Setup}
The simulation environment is based on site-specific maps of the University of Southern California (USC) campus, covering an area of $512 \times 512$ [m$^2$]. This environment reflects dense urban features, including buildings and terrain, providing a realistic scenario to evaluate AutoBS.

Simulations are executed on a high-performance machine with a Tesla T4 GPU, 12GiB of RAM, and an Intel Xeon CPU @ 2.20GHz. Developed in Python with PyTorch for model training, the framework leverages PMNet for pathloss predictions. 
% , significantly accelerating the reward computation of coverage and capacity over the site-specific map.

Table~\ref{table:paramter} in Appendix \ref{sec:appendixB} summarizes the key simulation parameters, including network configurations, environmental factors, and training setup.

%%%%%%%%%%%%%%%%%%%%%%%%%%%%
\BfPara{Benchmark}
To evaluate the performance of the proposed AutoBS framework, we compare it against two baseline methods: a heuristic-based deployment and an exhaustive search strategy. These baselines serve as the lower and upper bounds for performance comparison.

\begin{itemize}
    \item \textbf{Heuristic}:    
    This baseline simply places BSs uniformly at random within the deployable area $B$, with no learning or optimization. Such a random-placement approach serves as a performance lower bound since it ignores all site-specific information. (Notably, random spatial deployment is a common assumption in stochastic geometry models \cite{haenggi2012stochastic}.)

    \item \textbf{Exhaustive}:    
    % The exhaustive search provides an upper bound for performance by systematically evaluating all possible BS placement configurations across the site-specific map. It guarantees a global optimal solution in terms of coverage or capacity based on the configuration criteria. However, the computational cost is exceedingly high, particularly for multi-BS deployments, as the complexity grows exponentially with the number of BSs and map size. Due to this limitation, the exhaustive search is restricted to $50$ samples in our baseline experiments. Additionally, in multi-BS deployments, the exhaustive method operates synchronously, unlike the asynchronous deployment strategy used by AutoBS. While an asynchronous exhaustive method could be defined to reduce computation time somewhat, it would still be computationally intense and would yield suboptimal results. Our focus remains on synchronous exhaustive deployment as it provides a true upper bound for comparison with AutoBS.
    Exhaustive search evaluates every possible BS placement configuration on the map and thus provides a performance upper bound (it finds the globally optimal placement for a given metric). However, this approach is \emph{computationally prohibitive} – especially for multiple BSs, where complexity grows exponentially with the number of BSs and the map size. For practicality, we limit the exhaustive search in our experiments to evaluating 50 representative placements (rather than the full search space) as the multi-BS baseline. 
    % (In practice, we sample 50 candidate configurations because a true exhaustive evaluation is impossible for large scenarios.) 
    Additionally, our exhaustive baseline places multiple BSs \emph{synchronously} (optimizing all placements jointly), unlike AutoBS’s sequential addition. One could conceive of an “asynchronous” greedy exhaustive approach to speed it up, but it would still be extremely costly and not guarantee optimal results. We therefore use the synchronous exhaustive method as the true upper bound for comparison.

    \item \textbf{AutoBS (Proposed)}:    
    % AutoBS leverages the PPO agent to make near-optimal BS deployment decisions in real time. It is trained using a fully connected neural network with four layers, each consisting of 128 nodes, to optimize both coverage and capacity. AutoBS achieves rapid deployment decisions at millisecond timescales. 
    AutoBS uses a PPO-based agent to make near-optimal BS placement decisions in real time. The agent’s policy network is a fully-connected neural network with four hidden layers of 128 nodes each, trained to jointly optimize coverage and capacity. As a result, AutoBS can output deployment decisions on the order of only a few milliseconds.
    %, aided by the PMNet framework for fast and accurate pathloss predictions. With its focus on balancing performance and computational efficiency, AutoBS offers a practical, scalable solution suitable for large-scale network deployments.
\end{itemize}

%%%%%%%%%%%%%%%%%%%%%%%%%%%%
\subsection{Performance Analysis}

%%%%%%%%%%%%%%%%%%%%%%%%%%%%
\begin{table}[!h]
\centering
\resizebox{.35\linewidth}{!}{
\begin{minipage}{.45\linewidth}
% \small
\centering
\input{table/comparison_single-bs}

\end{minipage}}
\caption{Comparison results for Single (Static) BS deployment between exhaustive, random, and AutoBS deployment.}
\label{table:comparison_singlebs}
\vspace{-.5em}
\end{table}

\BfPara{Comparison}
Table \ref{table:comparison_singlebs} presents the performance of Heuristic, Exhaustive, and AutoBS deployments. AutoBS achieves significant improvements in both coverage and capacity compared to Heuristic deployment. Its performance closely approaches that of the Exhaustive method, which represents the global optimal solution, highlighting the effectiveness of policy-guided BS placement.

AutoBS focuses on coverage optimization through a reward function based on pathloss predictions from PMNet. Although capacity metrics are not explicitly included in the reward function (\ie $\nu_2=0$), the observed capacity difference between AutoBS and Exhaustive search remains minimal—which will be further discussed in Appendix~\ref{sec:appendixa}. 
%By balancing computational efficiency and network performance, AutoBS offers a scalable and practical solution for real-world, large-scale deployments.

%%%%%%%%%%%%%%%%%%%%%%%%%%%%
\begin{figure*}[!ht]
    \centering  
    \begin{subfigure}[t]{.22\linewidth}
        \centering  
        \includegraphics[width=\linewidth]{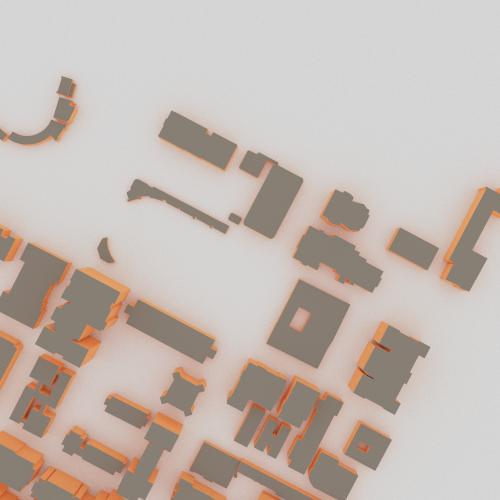}
        \caption{Building Map} \label{fig:bld map}
    \end{subfigure}\hspace*{.015\textwidth}%
    \begin{subfigure}[t]{.22\linewidth}
        \centering  
        \includegraphics[width=\linewidth]{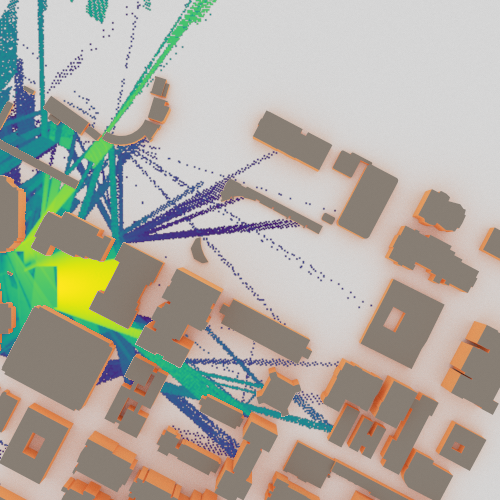}
        \caption{Heuristic}
    \end{subfigure}\hspace*{.015\textwidth}%
    \begin{subfigure}[t]{.22\linewidth}
        \centering
        \includegraphics[width=\linewidth]{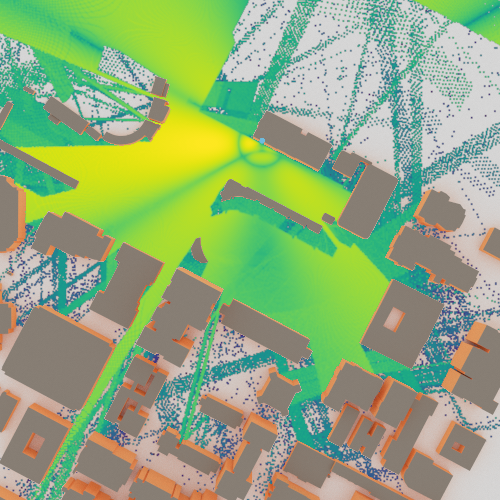}
        \caption{Exhaustive}
    \end{subfigure}
    \begin{subfigure}[t]{.22\linewidth}
        \centering  
        \includegraphics[width=\linewidth]{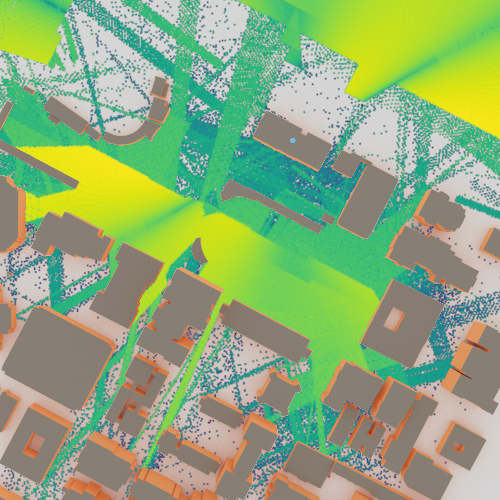}
        \caption{AutoBS}
    \end{subfigure}\hspace*{.015\textwidth}%
\caption{Coverage comparison for Single (Static) BS deployment. Simulations are performed using \emph{WirelessInsite} and visualized with \emph{SionnaRT}. Light green areas indicate regions with higher received signal strength. Fig. \ref{fig:bld map} illustrates the building map $\mathcal{S}$ used in each state $s_t$.}
\label{fig:comparison_coverage_singlebs}
\vspace{-.5em}
\end{figure*}

\BfPara{Coverage}
Fig. \ref{fig:comparison_coverage_singlebs} presents the coverage results for single BS deployment.\footnote{The plots are generated with SionnaRT purely because of its nicer graphical representation (the underlying numerical results are identical to those obtained with PMNet).} The building map, representing the input state, includes key geographical features such as buildings and obstacles, which significantly affect BS placement and signal propagation.

In the \emph{Heuristic} method, BSs are placed randomly across the deployable area, leading to inefficient coverage with noticeable gaps. This highlights the limitations of non-optimized deployments. The \emph{Exhaustive} search method evaluates every possible BS placement option, achieving the best coverage by systematically considering all configurations, but at the price of much higher complexity. 
%While this guarantees optimal results, it is computationally expensive and impractical for larger maps or multi-BS scenarios.

The \emph{AutoBS} method, leveraging the PPO algorithm and fast, accurate pathloss predictions from PMNet for efficient reward calculations, achieves coverage performance approaching that of exhaustive search, but with significantly reduced computational overhead.
%Although AutoBS doesn’t guarantee the absolute optimal solution, its learned policy effectively balances performance and efficiency.

% It is important to note that AutoBS uses PMNet during training for fast and accurate pathloss map generation.
%, while SionnaRT, which is more detailed but computationally demanding, is used only for visualization purposes and not for training \footnote{}.

\begin{figure}[!h]
\centering
\includegraphics[width=.35\linewidth]{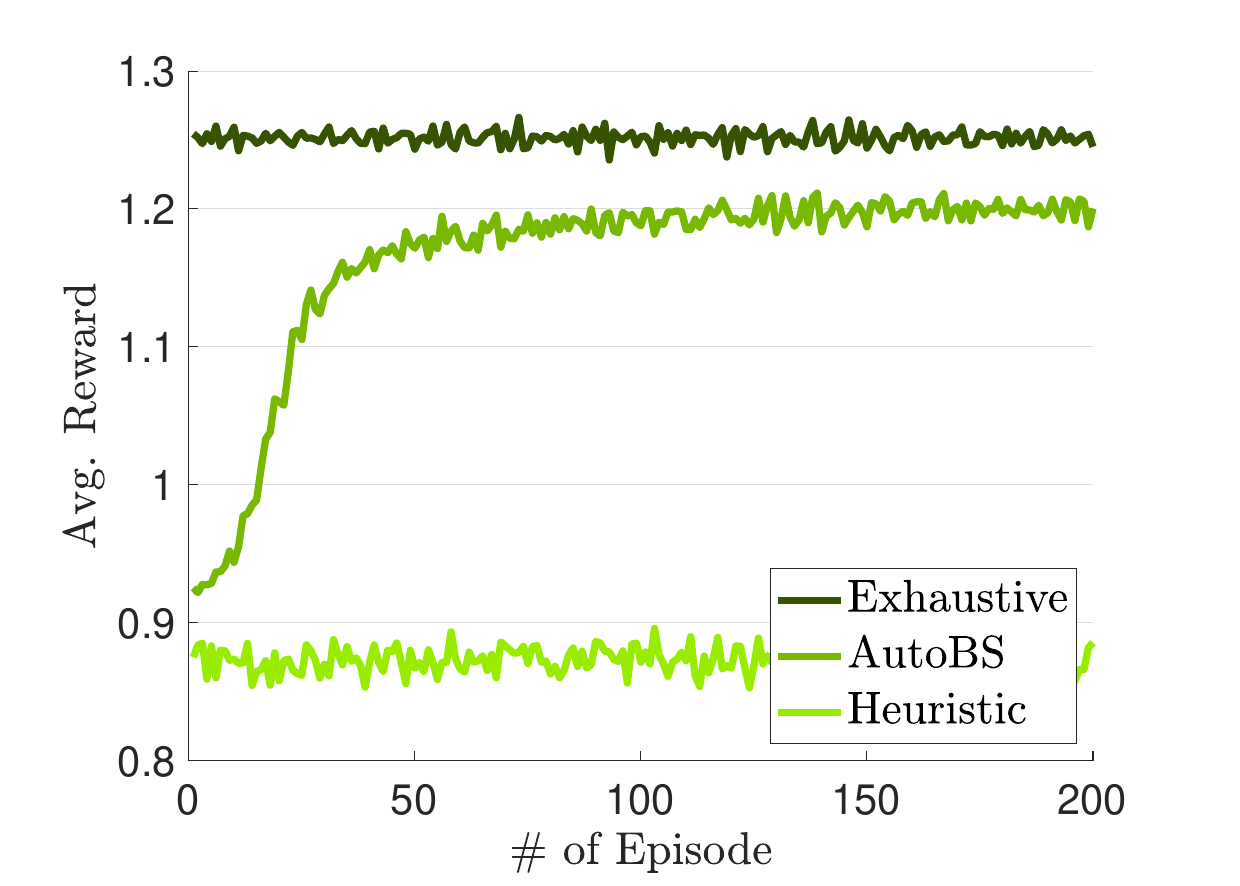}
\caption{Convergence behavior for Single (Static) BS deployment for Heuristic, Exhaustive, and AutoBS deployments over $200$ steps.}
\label{fig:convergence_singlebs}
\vspace{-.5em}
\end{figure}

%%%%%%%%%%%%%%%%%%%%%%%%%%%%
\BfPara{Convergence}
% Fig. \ref{fig:convergence_singlebs} presents the convergence behavior for single BS deployment, comparing AutoBS with Heuristic and Exhaustive methods (note that Heuristic and Exhaustive methods do not involve iterative learning). AutoBS stabilizes between $50$ to $100$ training episodes, approaching the performance of the global optimal solution (\ie Exhaustive), demonstrating its efficient learning and stable convergence.
Fig. \ref{fig:convergence_singlebs} shows the training convergence for the single-BS deployment scenario, comparing AutoBS with the baseline methods. (Note that the Heuristic and Exhaustive approaches do not involve iterative learning, so their performance is flat.) 
% We see that AutoBS’s performance steadily improves and stabilizes after approximately 50–100 training episodes, eventually approaching the level of the global optimum (Exhaustive). This demonstrates efficient learning and stable convergence of AutoBS agent.
AutoBS’s performance improves steadily and stabilizes after approximately 50–100 training episodes. Although a performance gap remains compared to the global optimum (Exhaustive), AutoBS achieves close-to-optimal performance efficiently, demonstrating effective learning and stable convergence.

%It is important to note that the Heuristic and Exhaustive methods do not involve iterative learning. The Heuristic deployment serves as a baseline, reflecting non-optimized lower-bound performance, while the Exhaustive deployment represents the global optimum but at the cost of high computational complexity. AutoBS offers a practical balance between optimality and computational efficiency. 
% Further quantitative comparisons of coverage and capacity can be found in Table~\ref{table:comparison_singlebs}.

\begin{figure*}[h!]
    \centering  
    \begin{subfigure}[t]{.22\linewidth}
        \centering  
        \includegraphics[width=\linewidth]{figure/building_map.png}
        \caption{Building map} \label{fig:bld map2}
    \end{subfigure}\hspace*{.015\textwidth}%
    \begin{subfigure}[t]{.22\linewidth}
        \centering  
        \includegraphics[width=\linewidth]{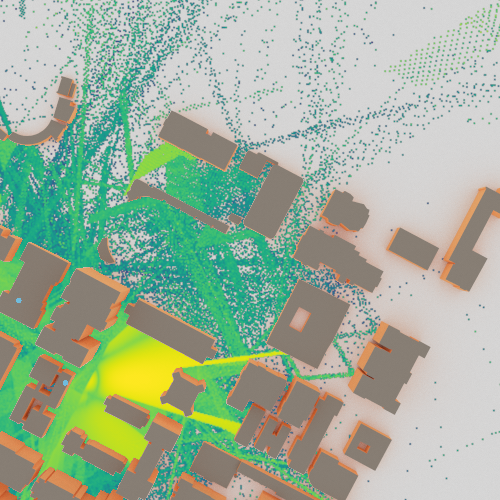}
        \caption{Heuristic}
    \end{subfigure}\hspace*{.015\textwidth}%
    \begin{subfigure}[t]{.22\linewidth}
        \centering
        \includegraphics[width=\linewidth]{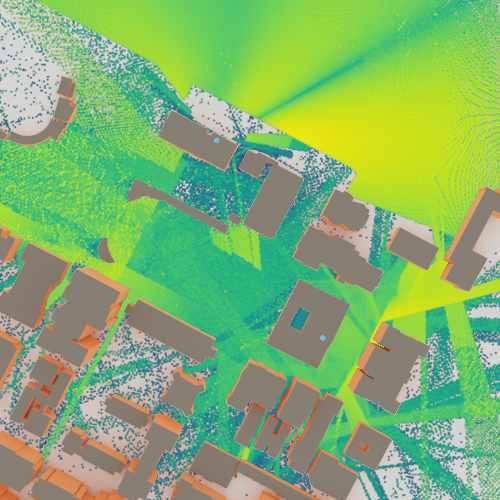}
        \caption{Exhaustive}
    \end{subfigure}
    \begin{subfigure}[t]{.22\linewidth}
        \centering  
        \includegraphics[width=\linewidth]{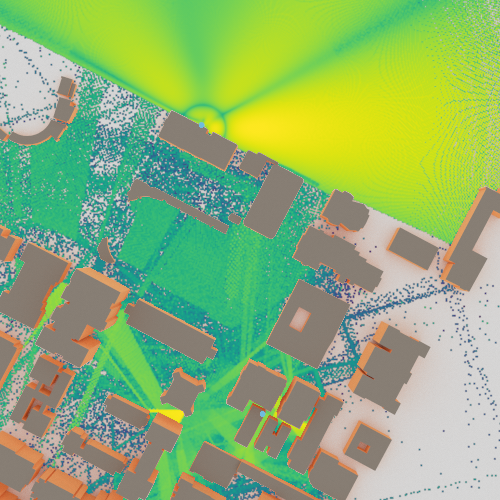}
        \caption{AutoBS}
    \end{subfigure}\hspace*{.015\textwidth}%
\caption{Comparison results for Multi (Asynchronous) BS deployment in terms of coverage using \emph{SionnaRT}. Light green areas indicate higher received signal strength. Fig. \ref{fig:bld map2} shows a building map $\mathcal{S}$ used in each state $s_t$.}
\label{fig:comparison_coverage_multibs}
\vspace{-.5em}
\end{figure*}

%%%%%%%%%%%%%%%%%%%%%%%%%%%%
\BfPara{Multi-BS Deployment (Asynchronous)}
% Table \ref{table:comparison_multibs} and Fig. \ref{fig:comparison_coverage_multibs} present the results for asynchronous multi-BS deployment. 
% The reward function is computed based only on the specific metrics (e.g., coverage, capacity, and/or pathgain) of each newly deployed BS. For example, when deploying the second BS, only the performance metrics for that BS are considered in the reward calculation.
Table \ref{table:comparison_multibs} and Fig. \ref{fig:comparison_coverage_multibs} summarize the results for the asynchronous multi-BS scenario. Note: In this setting, the reward at each step accounts only for the newly deployed BS’s contribution (\eg the coverage, capacity, and/or path gain added by that BS). For example, when deploying a second BS, only the second BS’s performance metrics are used to calculate the reward for that step.

\begin{table}[!h]
% \vspace{-.5em}
\centering
\resizebox{.35\linewidth}{!}{\begin{minipage}{.45\linewidth}
% \small
\centering
\input{table/comparison_multi-bs}

\end{minipage}}
\caption{Comparison results for Multi (Asynchronous) BS deployment between exhaustive, random, and AutoBS deployment.}
\label{table:comparison_multibs}
\vspace{-.5em}
\end{table}

% For the multi-BS case, AutoBS performs an asynchronous deployment, \ie adds one BS at a time,  in a "greedy" manner. This can be a practical requirement, \eg when adding BSs to an existing infrastructure. In a greenfield deployment (like investigated here), the sequential approach will give worse results than a synchronous approach. Potential iterations to fine-tune already-placed BSs will be subject of our future research.  
In our multi-BS experiments, AutoBS follows an asynchronous (greedy) deployment strategy, adding one BS at a time. This sequential approach is practical for scenarios like gradually adding BSs to an existing network. However, in a greenfield deployment (with no initial infrastructure, as studied here), a purely sequential addition can yield worse results than an ideal joint (synchronous) optimization. Developing methods to iteratively fine-tune or reposition previously placed BSs could mitigate this, and will be addressed in future work.

% While the overall findings align with those from the single (static) BS deployment, some notable insights emerge. One key observation is that deploying two BSs using the Heuristic method performs similarly or even worse than deploying a single BS with AutoBS or Exhaustive search in terms of coverage and capacity. This emphasizes the critical role of optimized deployment strategies. As shown in Table \ref{table:comparison_multibs} for Heuristic and Table \ref{table:comparison_singlebs} for AutoBS and Exhaustive search, non-optimized deployments fall short even when more BSs are deployed.
Overall, the multi-BS results align with the single-BS findings, but with a notable insight: using two randomly placed BSs (Heuristic) can perform no better – or even worse – than using just one well-placed BS (as AutoBS or Exhaustive would). In other words, simply deploying more BSs does not guarantee better performance without optimization. This highlights the critical importance of optimized deployment strategies. As seen by comparing Table~\ref{table:comparison_multibs} (Heuristic two-BS) with Table~\ref{table:comparison_singlebs} (AutoBS/Exhaustive single-BS), a non-optimized approach can fall short even with additional infrastructure.

% These findings highlight two key aspects. First, AutoBS, despite using an asynchronous deployment strategy, achieves $90$\% of the capacity of the global optimum (\ie provided by the exhaustive method). Second, AutoBS’s ability to learn site-specific characteristics enables it to efficiently deploy BSs. 
In summary, our multi-BS results highlight two key points: (1) AutoBS’s sequential (asynchronous) strategy still achieves roughly $90$\% of the capacity of the global optimum (the exhaustive placement), and (2) by learning the site-specific propagation characteristics, AutoBS is able to deploy multiple BSs very efficiently (vastly outperforming unoptimized placements).

% The results demonstrate that AutoBS offers a scalable, practical alternative to exhaustive search, achieving competitive performance without the prohibitive computational costs—which will be elaborated in the following.
% Another key insight is the minimal convergence gap between AutoBS and Exhaustive search, especially in the multi-BS scenario, as illustrated in Fig. \ref{fig:convergence_multibs}. 

% \begin{figure}[!h]
% \centering
% \includegraphics[width=1.\linewidth]{figure/convergence/v35.png}
% \caption{Convergence behavior for Multi (Asynchronous) BS deployment for Heuristic, Exhaustive, and AutoBS deployments over $200$ steps.}
% \label{fig:convergence_multibs}
% % \vspace{-.5em}
% \end{figure}

%%%%%%%%%%%%%%%%%%%%%%%%%%%%
\subsection{Complexity Analysis}
% Table \ref{table:complexity} highlights the time complexity analysis, emphasizing a key advantage of AutoBS: its ability to drastically reduce inference time, achieving millisecond-level decisions even in multi-BS deployments. 
% %AutoBS efficiently determines near-optimal deployment locations with minimal computational overhead, presenting high performance in both single and multi-BS scenarios.

% The efficiency gains are particularly notable in multi-BS deployments, where exhaustive search faces exponential complexity. AutoBS’s asynchronous deployment strategy enables it to scale seamlessly with growing network size, making it well-suited for large-scale networks and real-time optimization. 
% % By balancing computational efficiency with near-optimal performance, AutoBS offers a practical, scalable solution for network operators to manage complex deployment challenges in dynamic environments.

Table \ref{table:complexity} compares the computation time required by AutoBS versus an exhaustive search. The key takeaway is that AutoBS drastically reduces inference time — achieving decisions in milliseconds even for multi-BS deployments — whereas exhaustive search quickly becomes intractable as the number of BSs grows. The efficiency gains of AutoBS are especially dramatic in the multi-BS case (exhaustive search time explodes exponentially with each additional BS). AutoBS’s sequential deployment strategy, on the other hand, scales easily with network size, demonstrating that it is well-suited for large networks and real-time optimization.

\begin{table*}[h!]
% \vspace{-.5em}
\centering
\resizebox{.7\linewidth}{!}{\begin{minipage}{.73\linewidth}
\small
\input{table/complexity}
\end{minipage}}
\caption{Time complexity analysis.}
\label{table:complexity}
\vspace{-2.em}
\end{table*}

%% file: table/comparison_single-bs.tex
\begin{tabularx}{1\linewidth}{l c c}
\toprule[1pt]
\textbf{\textit{Scheme}} & \textbf{\textit{Coverage} ($V$)} & \textbf{\textit{Capacity} ($C$)}  \\
\midrule[.8pt]

Heuristic & $43.37 \%$ & $1.892 $ \\
AutoBS & $59.44 \%$ & $2.893 $ \\
\cmidrule(lr){1-3}
Exhaustive for $V$ & $64.23 \%$ & $2.981$ \\
Exhaustive for $C$ & $63.08 \%$ & $3.036$  \\

\bottomrule[1pt]
\end{tabularx}

%% file: table/comparison_multi-bs.tex
\begin{tabularx}{1\linewidth}{l c c}
\toprule[1pt]
\textbf{\textit{Scheme}} & \textbf{\textit{Coverage} ($V$)} & \textbf{\textit{Capacity} ($C$)}  \\
\midrule[.8pt]

Heuristic & $63.95 \%$ & $2.966$ \\
AutoBS & $71.65 \%$ & $3.637$ \\
\cmidrule(lr){1-3}
Exhaustive for $V$ & $81.06 \%$ & $3.995$ \\
Exhaustive for $C$ & $75.36\%$ & $4.023$ \\

\bottomrule[1pt]
\end{tabularx}

%% file: table/complexity.tex
% \begin{tabularx}{\linewidth}{l l}
% \toprule[1pt]
% \textbf{\textit{Method}} & \textbf{\textit{Inference Time}} [sec] \\
% \midrule[.8pt]

% \emph{\textbf{Single BS Deploy.}} & \\
% \cmidrule(lr){1-1} \cmidrule(lr){2-2}
% Exhaustive & $3.2245$ ($\nicefrac{10000}{10000}$) \\
% AutoBS & $0.0145$ ($\nicefrac{45}{10000}$) \\

% \cmidrule(lr){1-2}

% \emph{\textbf{Two BS Deploy.}} & \\
% \cmidrule(lr){1-1} \cmidrule(lr){2-2}
% Exhaustive & $4062.1$ ($\nicefrac{1000000}{1000000}$) \\
% % Exhaustive (Async.) & $6.4491$ ($\nicefrac{xXXX}{1000000}$) \\
% AutoBS & $0.1608$ ($\nicefrac{39}{1000000}$) \\

% \cmidrule(lr){1-2}

% \emph{\textbf{Three BS Deploy.}} & \\
% \cmidrule(lr){1-1} \cmidrule(lr){2-2}
% Exhaustive & $13449$ ($\nicefrac{1000000}{1000000}$) \\
% AutoBS & $0.2128$ ($\nicefrac{15}{1000000}$) \\

% \bottomrule[1pt]
% \end{tabularx}

\begin{tabularx}{\linewidth}{l l l l}
\toprule[1pt]
\textbf{\textit{}} & \textbf{\textit{One BS Deploy.}} [sec] & \textbf{\textit{Two BS Deploy.}} [sec] & \textbf{\textit{Three BS Deploy.}} [sec] \\
\midrule[.8pt]
Exhaustive & 3.2245 ($\nicefrac{10000}{10000}$) & 4062.1 ($\nicefrac{1000000}{1000000}$) & 13449 ($\nicefrac{1000000}{1000000}$) \\
AutoBS & \textbf{0.0145} ($\nicefrac{45}{10000}$) & \textbf{0.1608} ($\nicefrac{39}{1000000}$) & \textbf{0.2128} ($\nicefrac{15}{1000000}$) \\
\bottomrule[1pt]
\end{tabularx}

%% file: section/conclusion.tex
\section{Conclusion} \label{sec:conclusion}
In this paper we presented AutoBS, a reinforcement-learning-based framework for optimal BS deployment in 6G RANs. By leveraging the PPO algorithm and integrating fast, accurate pathloss predictions from PMNet (a generative model for DNT), AutoBS efficiently learns deployment strategies that adapt to site-specific channel conditions while balancing coverage and capacity.

Our results show that AutoBS achieves \emph{near-optimal} performance – up to $\sim95 \%$ of the capacity of exhaustive search – while \emph{reducing inference time from hours to milliseconds}. This makes AutoBS a practical solution for real-time deployment. With its combination of near-optimal performance and efficient scalability, AutoBS stands out as a powerful automated solution tailored to the demands of large-scale 6G network optimization.

While we demonstrated AutoBS for 6G BS deployment, the underlying framework can be easily adapted to other network optimization tasks in DNT setting. By modifying the state representation and reward function, the same approach could tackle problems like mobility management, link adaptation, or BS sleep mode energy savings, aligning the optimization with specific application goals. Furthermore, AutoBS’s flexible design makes it suitable for entirely different deployment scenarios – for example, optimal placement of Wi-Fi access points (APs) or O-RAN radio units (RUs).

%% file: section/appendix.tex
\section{Reward Design} \label{sec:appendixa}

% \vspace{-.5em}
\begin{table}[h!]
\centering
\resizebox{.55\linewidth}{!}{
\begin{minipage}{.5\linewidth}
\centering
% \small
\input{table/comparison_single-bs_custom}

\end{minipage}}
\caption{Reward design variations for single (static) BS deployment.}
\label{table:reward}
\vspace{.5em}
\end{table}

This section presents empirical findings from experiments with different reward designs in AutoBS framework. The pathgain is defined as $P = \sum_{{i,j} \in \mathcal{R}}{P_{i,j}}$, and Table \ref{table:reward} summarizes coverage and capacity outcomes for each reward configuration. Although other transformations, such as logarithmic scaling (\eg $\log{V}$), were tested, they are omitted here for brevity. These findings are empirical and sensitive to training hyperparameters (\eg learning rate) and reward design parameters (\eg $\nu$).

The results provide valuable insights. As expected, the Coverage Only reward achieves the highest coverage, aligning with its direct optimization objective (\eg $\nu_2, \nu_3 = 0$). Interestingly, in terms of capacity, the combination of Pathgain and Coverage (\ie $\nu_2=0$) outperforms the Capacity Only reward. Notably, Coverage Only also surpasses Capacity Only in capacity, highlighting complex dynamics within the DRL training process. These findings suggest that site-specific channel characteristics are implicitly embedded in rewards derived from PMNet’s pathloss maps. While Capacity Only smooths variations through logarithmic scaling, Pathgain captures these fluctuations on a linear scale, providing more granular feedback for effective learning.

Throughout this work, the Pathgain + Coverage reward configuration is primarily used.

\newpage
\section{Setup} \label{sec:appendixB}
For all experiments, we used the same simulation configuration. Table~\ref{table:paramter} below summarizes the key parameters of this setup, divided into two parts: the wireless network environment settings and the training hyperparameters for AutoBS.

\begin{table}[!h]
\centering
\resizebox{.6\columnwidth}{!}{
% \begin{minipage}[t]{.9\columnwidth}
\centering
\input{table/parameters_extended}
}
\caption{Simulation Parameters.}
\label{table:paramter}
\end{table}

%% file: table/comparison_single-bs_custom.tex
\begin{tabularx}{1\linewidth}{l c c}
\toprule[1pt]
\textbf{\textit{Reward}} & \textbf{\textit{Coverage} ($V$)} & \textbf{\textit{Capacity} ($C$)}  \\
\midrule[.8pt]

% Exponential-Reward & $45.125 \%$ \; \tikz{
% \draw[gray,line width=.3pt] (0,0) -- (1.1,0);
% \draw[white, line width=0.01pt] (0,-2pt) -- (0,2pt);
% \draw[black,line width=1pt] (0.14,0) -- (0.3,0);
% \draw[black,line width=1pt] (0.14,-2pt) -- (0.14,2pt);
% \draw[black,line width=1pt] (0.3,-2pt) -- (0.3,2pt);} 
% & $2.214$  \; \tikz{
% \draw[gray,line width=.3pt] (0,0) -- (1.1,0);
% \draw[white, line width=0.01pt] (0,-2pt) -- (0,2pt);
% \draw[black,line width=1pt] (0.05,0) -- (0.15,0);
% \draw[black,line width=1pt] (0.05,-2pt) -- (0.05,2pt);
% \draw[black,line width=1pt] (0.15,-2pt) -- (0.15,2pt);}
% \\

% Relu-Reward  & $43.707 \%$ \; \tikz{
% \draw[gray,line width=.3pt] (0,0) -- (1.1,0);
% \draw[white, line width=0.01pt] (0,-2pt) -- (0,2pt);
% \draw[black,line width=1pt] (0.05,0) -- (0.15,0);
% \draw[black,line width=1pt] (0.05,-2pt) -- (0.05,2pt);
% \draw[black,line width=1pt] (0.15,-2pt) -- (0.15,2pt);}
% & $2.094$  \; \tikz{
% \draw[gray,line width=.3pt] (0,0) -- (1.1,0);
% \draw[white, line width=0.01pt] (0,-2pt) -- (0,2pt);
% \draw[black,line width=1pt] (0.05,0) -- (0.15,0);
% \draw[black,line width=1pt] (0.05,-2pt) -- (0.05,2pt);
% \draw[black,line width=1pt] (0.15,-2pt) -- (0.15,2pt);}
% \\

Capacity Only  & $57.642 \%$ 
% \; \tikz{
% \draw[gray,line width=.3pt] (0,0) -- (1.1,0);
% \draw[white, line width=0.01pt] (0,-2pt) -- (0,2pt);
% \draw[black,line width=1pt] (0.05,0) -- (0.15,0);
% \draw[black,line width=1pt] (0.05,-2pt) -- (0.05,2pt);
% \draw[black,line width=1pt] (0.15,-2pt) -- (0.15,2pt);}
& $2.862$  
% \; \tikz{
% \draw[gray,line width=.3pt] (0,0) -- (1.1,0);
% \draw[white, line width=0.01pt] (0,-2pt) -- (0,2pt);
% \draw[black,line width=1pt] (0.05,0) -- (0.15,0);
% \draw[black,line width=1pt] (0.05,-2pt) -- (0.05,2pt);
% \draw[black,line width=1pt] (0.15,-2pt) -- (0.15,2pt);}
\\

Coverage + Capacity  & $57.833 \%$ 
% \; \tikz{
% \draw[gray,line width=.3pt] (0,0) -- (1.1,0);
% \draw[white, line width=0.01pt] (0,-2pt) -- (0,2pt);
% \draw[black,line width=1pt] (0.05,0) -- (0.15,0);
% \draw[black,line width=1pt] (0.05,-2pt) -- (0.05,2pt);
% \draw[black,line width=1pt] (0.15,-2pt) -- (0.15,2pt);}
& $2.831$  
% \; \tikz{
% \draw[gray,line width=.3pt] (0,0) -- (1.1,0);
% \draw[white, line width=0.01pt] (0,-2pt) -- (0,2pt);
% \draw[black,line width=1pt] (0.05,0) -- (0.15,0);
% \draw[black,line width=1pt] (0.05,-2pt) -- (0.05,2pt);
% \draw[black,line width=1pt] (0.15,-2pt) -- (0.15,2pt);}
\\

Pathgain + Capacity  & $58.747 \%$ 
% \; \tikz{
% \draw[gray,line width=.3pt] (0,0) -- (1.1,0);
% \draw[white, line width=0.01pt] (0,-2pt) -- (0,2pt);
% \draw[black,line width=1pt] (0.05,0) -- (0.15,0);
% \draw[black,line width=1pt] (0.05,-2pt) -- (0.05,2pt);
% \draw[black,line width=1pt] (0.15,-2pt) -- (0.15,2pt);}
& $2.883$  
% \; \tikz{
% \draw[gray,line width=.3pt] (0,0) -- (1.1,0);
% \draw[white, line width=0.01pt] (0,-2pt) -- (0,2pt);
% \draw[black,line width=1pt] (0.05,0) -- (0.15,0);
% \draw[black,line width=1pt] (0.05,-2pt) -- (0.05,2pt);
% \draw[black,line width=1pt] (0.15,-2pt) -- (0.15,2pt);}
\\

Coverage Only & $\textbf{59.537} \% $ 
% \; \tikz{
% \draw[gray,line width=.3pt] (0,0) -- (1.1,0);
% \draw[white, line width=0.01pt] (0,-2pt) -- (0,2pt);
% \draw[black,line width=1pt] (0.14,0) -- (0.3,0);
% \draw[black,line width=1pt] (0.14,-2pt) -- (0.14,2pt);
% \draw[black,line width=1pt] (0.3,-2pt) -- (0.3,2pt);} 
& $2.884$  
% \; \tikz{
% \draw[gray,line width=.3pt] (0,0) -- (1.1,0);
% \draw[white, line width=0.01pt] (0,-2pt) -- (0,2pt);
% \draw[black,line width=1pt] (0.05,0) -- (0.15,0);
% \draw[black,line width=1pt] (0.05,-2pt) -- (0.05,2pt);
% \draw[black,line width=1pt] (0.15,-2pt) -- (0.15,2pt);}
\\

Pathgain + Coverage & $59.438 \%$ 
% \; \tikz{
% \draw[gray,line width=.3pt] (0,0) -- (1.1,0);
% \draw[white, line width=0.01pt] (0,-2pt) -- (0,2pt);
% \draw[black,line width=1pt] (0.05,0) -- (0.15,0);
% \draw[black,line width=1pt] (0.05,-2pt) -- (0.05,2pt);
% \draw[black,line width=1pt] (0.15,-2pt) -- (0.15,2pt);}
& $\textbf{2.893}$  
% \; \tikz{
% \draw[gray,line width=.3pt] (0,0) -- (1.1,0);
% \draw[white, line width=0.01pt] (0,-2pt) -- (0,2pt);
% \draw[black,line width=1pt] (0.05,0) -- (0.15,0);
% \draw[black,line width=1pt] (0.05,-2pt) -- (0.05,2pt);
% \draw[black,line width=1pt] (0.15,-2pt) -- (0.15,2pt);}
\\

\bottomrule[1pt]
\end{tabularx}

%% file: table/parameters_extended.tex
\begin{tabular} {l l}
\toprule[1pt]
% \textbf{\textit{Type (or Parameter)}} & \textbf{\textit{Value}} \\
% \midrule[.8pt]

\textbf{Network Config.} \\
\cmidrule(lr){1-1} \cmidrule(lr){2-2}
Area & $512 \times 512$ [m$^2$]  \\ 
Carrier frequency & $2.5$ [GHz] \\
Effective bandwidth & $1$ [MHz] \\
Antenna & Isotropic (vertical) \\
Input power & $0$ [dBm] \\
Noise figure & $3$ [dB] \\
Coverage threshold & $-90.015$ [dBm]  \\

\cmidrule(lr){1-2}
\textbf{AutoBS Train Config.} \\
\cmidrule(lr){1-1} \cmidrule(lr){2-2}

Generative digital twin model & PMNet$_{\mathrm{usc}}$ \cite{lee2024scalable} \\ 
DRL algorithm & PPO \cite{schulman2017proximal} \\ 
Reward function & Pathgain + Coverage ($\nu_2=0$) \\ 

Learning rate & $1.0 \times 10^{-5}$ \\
Gamma & $0.1$ \\
% Gradient Clipping & $40.0$ \\
% Exploration type Size & StochasticSampling \\
SGD minibatch size & $256$ \\

\# of samples for train (test) & $1100$ ($50$)\\
\# of episode & $5$ 
\\

\bottomrule[1pt]
\end{tabular}